# Molecular Recognition as an Information Channel: The Role of Conformational Changes

Yonatan Savir and Tsvi Tlusty

Dept. of Physics of Complex Systems
The Weizmann Institute of Science
Rehovot 76100, Israel
Email: yonatan.savir@weizmann.ac.il

***Abstract*—Molecular recognition, which is essential in processing information in biological systems, takes place in a crowded noisy biochemical environment and requires the recognition of a specific target within a background of various similar competing molecules. We consider molecular recognition as a transmission of information via a noisy channel and use this analogy to gain insights on the optimal, or fittest, molecular recognizer. We focus on the optimal structural properties of the molecules such as flexibility and conformation. We show that conformational changes upon binding, which often occur during molecular recognition, may optimize the detection performance of the recognizer. We thus suggest a generic design principle termed 'conformational proofreading' in which deformation enhances detection. We evaluate the optimal flexibility of the molecular recognizer, which is analogous to the stochasticity in a decision unit. In some scenarios, a flexible recognizer, i.e., a stochastic decision unit, performs better than a rigid, deterministic one. As a biological example, we discuss conformational changes during homologous recombination, the process of genetic exchange between two DNA strands.**

## I. INTRODUCTION

Molecular recognition plays a key role in processing information in biological systems. Processes such as decoding DNA sequences by regulatory proteins, identification of antigens by antibodies and signal transduction by enzymes lean on the ability of molecules to recognize and bind a *specific* target. However, the crowded biological environment contains many molecules with similar structure that may compete with the "right" target. Moreover, recognition is often carried out using non-covalent binding energies that are of the same order as the stochastic thermal energy [1]. Thus, the recognition process is prone to false binding, which introduces errors and may impair the proper information flow. As a result, noise must be taken into account in the design and the evolution of molecular information channels and, specifically, molecular codes [2, 3].

The remarkable efficiency and specificity of molecular recognition led Emil Fisher already in 1890 to propose the 'Lock and Key' model which postulates that an enzyme and a substrate should be complementary in shape and thus discriminate against other substrates that do not fit to enzyme binding site [Fig. 1(a)]. Later, studies indicated that the 'native' conformations of many molecular recognizers and targets may not be exactly complementary ('Induced fit' [4]). Therefore, they may deform in order to bind each other and conformational changes occur upon binding [5-9].

Unlike the picture that arises from the 'Lock and Key' model, the interacting molecules are not always rigid objects that are complementary in structure. In the noisy biological environment, the molecular recognition process may involve molecules that fluctuate around non-complementary native conformations. As a result, a variety of complexes may be formed upon binding [Fig. 1(b)]. This leads to the question of whether conformational changes upon binding are merely a biochemical constrain or whether they are the outcome of an evolutionary pressure to optimize molecular recognition.

In this work, we draw the analogies between molecular recognition and transmission of information via a noisy channel and discuss the optimal design of a molecular recognizer [10, 11]. Our analysis show that the optimal design depends on the flexibility of the molecules. For typical biological values of flexibility, conformational changes upon binding may optimize the detection performance of the recognizer. The flexibility of the recognizer is analogous to stochasticity in the decision unit. Soft molecules with high flexibility fluctuate more than molecules with lower flexibility and thus are more 'noisy'. In some cases a flexible recognizer, a stochastic decision unit, performs better than a rigid, deterministic, one. As a biological case study, we discuss conformational changes during homologous recombination, an all-important process that facilitates the exchange of genetic material between homologous DNA molecules and their possible impact on correct detection [12].

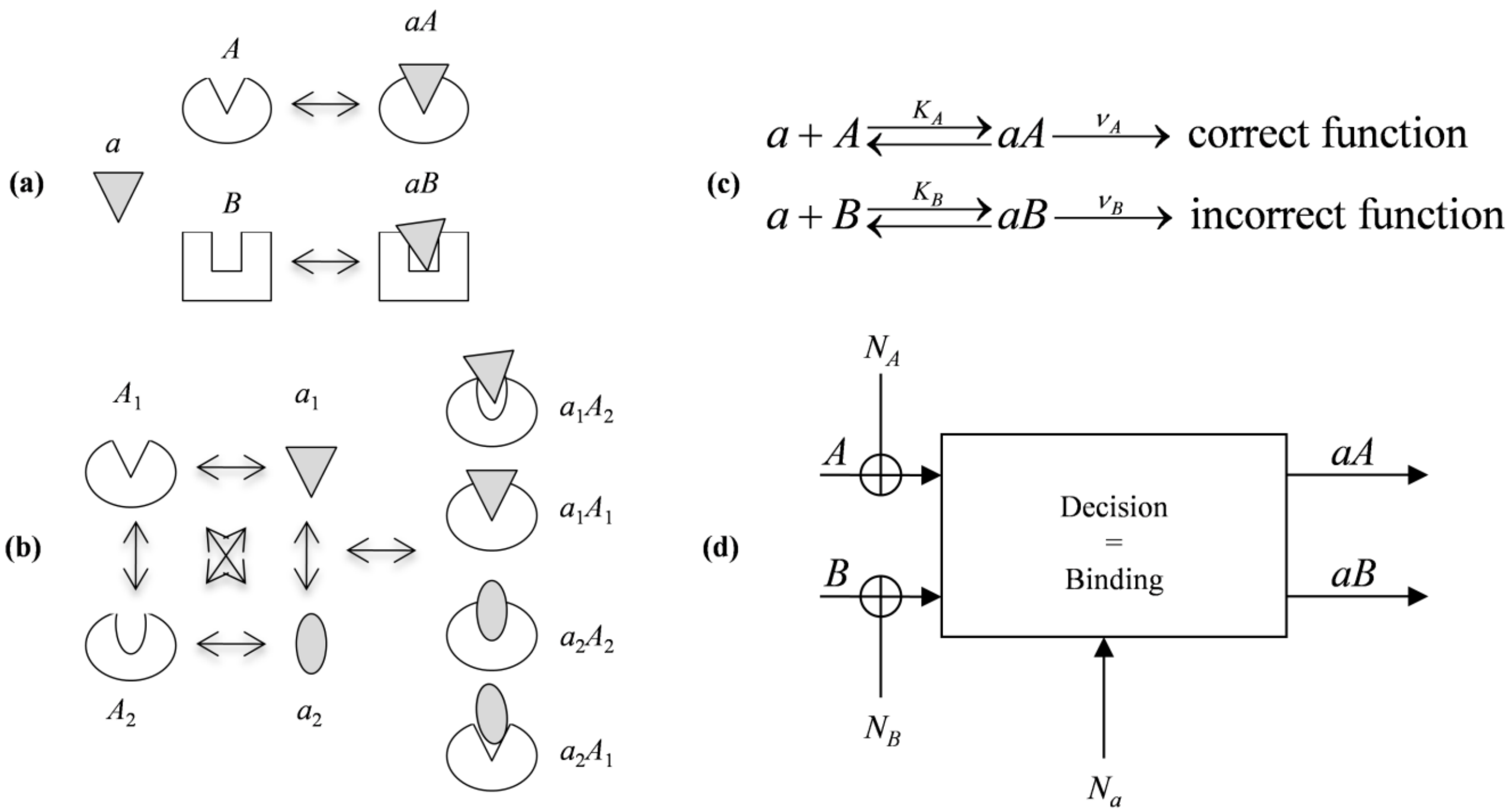

Fig. 1. Models of molecular recognition. (a) The 'Lock and Key' model. The bio-recognizer *a* has to discriminate between two competing targets *A* and *B*. The binding sites of the recognizer and its correct target *A* are complementary. Thus, their binding is tighter relative to the binding of the recognizer with an incorrect target *B* that has a non-complementary binding site. (b) Conformational changes upon binding. Both the recognizer *a* and its target *A* are fluctuating within an ensemble of conformations (only two are shown in the figure) and as a result a variety of complexes may be formed upon binding. (c) The typical recognition reaction can be described using Michaelis-Menten kinetics. In the first reversible step, the recognizer collides with a target and may bind it to form a complex. The stability of the formed complexes, *aA* or *aB*, depends on the dissociation constants, $K_A = [a][A]/[aA]$, $K_B = [a][B]/[aB]$ where [] denotes concentration. In the second, irreversible step, the bound complexes initiate some function with a rate $v_A$, in the case of *aA*, and with a rate $v_B$ in the case of *aB*. (d) Formulation of molecular recognition as a signal detection problem. The input signals are the target molecules *A* and *B* while the output is whether *A* or *B* are detected. In this analogy, the detection unit is the recognizer *a* and molecular binding governs the detection efficiency. Since all the molecules fluctuate within an ensemble of conformations, a noise should be added to the input signals and to the detection unit.

## II. MOLECULAR RECOGNITION AS A SIGNAL DETECTION PROBLEM

Let us consider a recognition process in which a recognizer *a*, for example an antibody, has to discriminate between two competing targets, *A*, a harmful pathogen, and *B*, a harmless self molecule [Fig. 1(c)]. The goal of the recognizer in this case is to trigger an immune response when it recognizes *A*. However, binding the harmless molecule *B* may lead to an undesirable auto-immune response. Molecular recognition processes typically follow Michaelis-Menten kinetics [13]. In the first reversible step, the recognizer collides with a target and may bind it to form a complex. The stability of the formed complexes, *aA* or *aB* in our case, depends on the dissociation constants, $K_A = [a][A]/[aA]$, $K_B = [a][B]/[aB]$, where [] denotes concentration. The dissociation constant is related to the free energy difference between the unbound state (free molecules) and the bound state (complex) measured in units of $k_BT$ ($\Delta G > 0$), $K \sim exp(-\Delta G)$. A complex with low free energy has a large $\Delta G$ and a small $K$ and thus is more stable. In the second, irreversible step, the bound complexes initiate a response, in our example an immune response, with a rate $v_A$, in the case of *aA*, and with a rate $v_B$ in the case of *aB*.

This recognition process can be treated as a signal detection problem [Fig. 1(d)]. The input signals are the target molecules *A* and *B*, while the output is whether to trigger or not to trigger an immune response (Table I). In this analogy, the detection unit is the recognizer *a*. Molecular binding, affected by the physical constrains on the molecules, governs the detection efficiency. Since all the molecules fluctuate within an ensemble of conformations, a noise should be added to the input signals and to the detection unit. The flexibility of the molecules affects their stochastic fluctuations and the noise associated with them. For example, a rigid recognizer does not fluctuate and thus can 'sample' only one conformation in a deterministic fashion. A flexible recognizer interconverts between an ensemble of conformations and therefore samples the conformations of its target in a stochastic fashion. We employ detection theory in order to evaluate the optimal conformational mismatch between a recognizer and its target and the optimal flexibility of these molecules.

A standard measure for the quality of detection systems is the Bayesian decision rule [14]. This rule is obtained by minimizing the Bayesian cost function, $C_b$, using posteriori probabilities,

TABLE I

EXAMPLE FOR A DECISION TABLE OF MOLECULAR RECOGNTION. AN ANTIGEN $a$ HAS TO DISCRIMINATE BETWEEN A PATHOGEN $A$ AND A HARMLESS MOLECULE $B$. THE DECISION IS MADE BY BINDING BETWEEN THE ANTIGEN AND THE TAREGTS. AS THE ANTIGEN BINDS ONE OF THE TARGETS, AN IMMUNE RESPONSE MAY BE TRIGGERED. THE TABLE SPECIFIES THE POSSIBLE DECISIONS AND THEIR PROBABILITIES. $p_h$ IS THE PROBABILITY THAT THE ANITGEN ENCOUNTERS $A$ OR $B$ WHILE $p_d$ IS THE CONDITIONAL DECISION PROBABILITY GIVEN THE INPUT.

| Decision \ Input | Pathogen $A$ $P_h(A)$ | Harmless Molecule $B$ $P_h(B)$ |
|---|---|---|
| Trigger response (t) | True positive: $a$ binds $A$ $P_d(t\|A)$ | False positive: $a$ binds $B$ $P_d(t\|B)$ |
| Do not trigger Response (nt) | False negative $1-P_d(t\|A)$ | True negative $1-P_d(t\|B)$ |

$$C_b = \sum_{i,j} C_{ij} p_h(j) p_d(i \mid j)\,, \tag{1}$$

where $p_h(j)$ is the probability for an input $j$ to occur and $p_d(i|j)$ is the conditional probability for an output $i$ given the input was $j$ ($\Sigma_i\, p_d(i|j) = 1$). $C_{ij}$ is the cost assigned for such a decision and measures the impact of each decision on the system. In the case of molecular recognition, the inputs are the encounters of the recognizer with $A$ or $B$, denoted by sub-indices $A$ and $B$, and the decisions are to trigger or not to trigger the response associated with $A$, denoted by sub-indices $t$ and $nt$.

Minimizing the Bayesian cost function (1) amounts to minimizing [11]

$$C = -c_{correct} \cdot p_h(A) \cdot p_d(t \mid A) + c_{incorrect} \cdot p_h(B) \cdot p_d(t \mid B), \tag{2}$$

where $c_{correct} = C_{nt,A} - C_{t,A}$ and $c_{incorrect} = C_{t,B} - C_{nt,B}$. Since the weights of the correct decisions, $C_{t,A}$ and $C_{nt,B}$, are negative and the weights of the false decisions, $C_{t,B}$ and $C_{nt,A}$, are positive, both $c_{correct}$ and $c_{incorrect}$ are positive. Thus, increasing the conditional true positive probability $p_d(t|A)$, reduces the cost while increasing the conditional false positive probability $p_d(t|B)$, increases the cost. At the molecular level, $p_h$ is the probability that the recognizer encounters one of the targets. The conditional detection probability, $p_d$, is the product of the binding probability between the molecules and the probability that the formed complex is functional, $p_f$, $p_d = p_b \cdot p_f$. Therefore, the cost function (2) takes the form,

$$C = -c_{correct} \cdot p_h(A) \cdot p_b(A) \cdot p_f(aA) + c_{incorrect} \cdot p_h(B) \cdot p_b(B) \cdot p_f(aB). \tag{3}$$

In this form, the cost function has a clear biological meaning. The first term of (3) is proportional to the rate of correct function induction, whereas the second term is proportional to the rate of incorrect function induction.

The binding probability between the corresponding targets can be estimated by $p_b = 1/[1 + \exp(-\Delta G_t)]$, where $\Delta G_t$ is the total free energy difference between the unbound and bound states and is measured in units of $k_B T$ ($\Delta G_t > 0$). As the recognizer and its target come into close contact they may deform in order to align their binding sites [Fig. 1(b)]. This deformation requires the investment of free energy, $\Delta G_{def}$. Once the binding sites are aligned, binding free energy, $\Delta G_{int}$, is gained. The total free energy difference is thus, $\Delta G_t = \Delta G_{int} - \Delta G_{def}$.

By introducing the binding probability, $p_b$, into the detection cost $C$ (3), we obtain a measure for the quality of detection as a function of the interaction and deformation free energies,

$$C = \frac{-1}{1 + exp(\Delta G_{def} - \Delta G_{int,A})} + t \cdot \frac{1}{1 + exp(\Delta G_{def} - \Delta G_{int,B})}, \tag{4}$$

where the sub-indices $A$ and $B$ denote the free energies of the correct and incorrect targets, respectively. The binding interactions with the correct target are stronger than the binding interactions with the incorrect one, $\Delta G_{int,A} > \Delta G_{int,B}$. For clarity we have assumed that the binding sites of the competing targets are similar and thus the deformation free energy required to bind them is almost the same. The first term of (4) is the benefit from detecting the correct target $A$, increasing true positives and decreasing false negatives, whereas the second term is the penalty for detecting the incorrect target $B$, increasing false positives and decreasing true negatives. The parameter $t = [c_{incorrect} p_h(B) p_f(aB)]/[c_{correct} p_h(A) p_f(aA)]$ is the *tolerance* of the system. As $t$ increases, the penalty for detecting an incorrect target increases and the system is less tolerant of errors. For example, this may occur when the penalty for an incorrect decision, $c_{incorrect}$, is much higher than the benefit from a correct one, $c_{correct}$, or when incorrect targets are more abundant than correct ones. In a similar fashion, an error-tolerant system is characterized by a lower value of $t$.

The optimal deformation free energy is the outcome of a tradeoff between maximizing the probability of binding to the correct target, while minimizing the probability of binding to the incorrect one. If the optimal deformation free energy value is zero, no deformation is needed and the native states of the molecules should be complementary ('locks and key'). However, as we show below, if the optimal deformation free energy is not zero, the optimal recognizer should not be complementary to its target and thus conformational changes occur upon binding.

The binding probability as a function of the deformation free energy is a sharp sigmoid [Fig. 2(a)]. As the deformation free energy increases, the correct binding probability decreases. The correct binding probability reaches half its maximal value at $\Delta G_{def} \approx \Delta G_{int,A}$. At the same time, the probability of incorrect binding also decreases and reaches half its maximal value at $\Delta G_{def} \approx \Delta G_{int,B}$. The reduction of the incorrect binding probability is steeper than the correct binding probability. Therefore, for the symmetric case in which the impact of correct and incorrect decision is equal, $t$ = 1, the cost function exhibits a minimum at a non-zero value, $\Delta G_{def}$ = 1/2( $\Delta G_{int,A}$+ $\Delta G_{int,B}$).

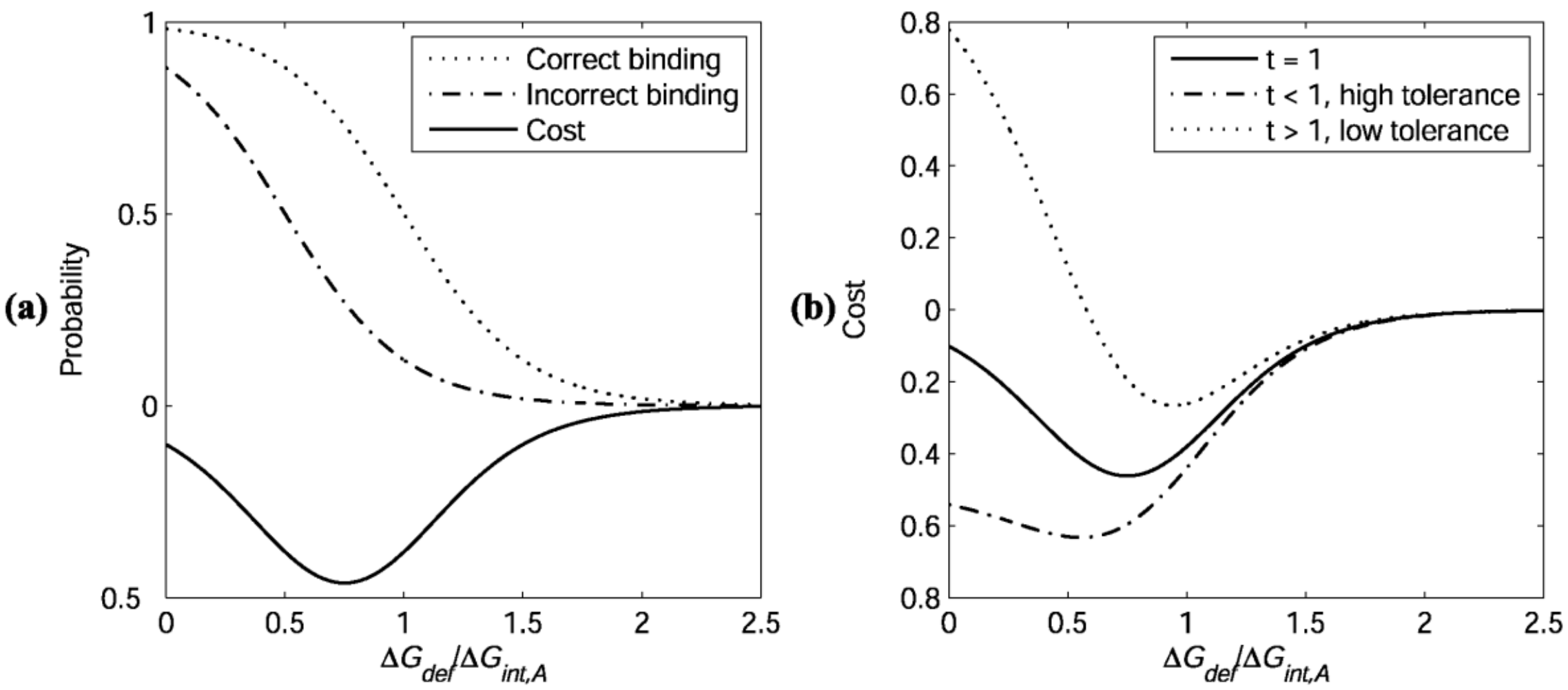

Fig. 2. (a) The cost function and its dependence on the deformation free energy for the symmetric case in which the impact of correct and incorrect decision is equal, $t$ = 1. $\Delta G_{int,A}$ and $\Delta G_{int,B}$ are the free energy gains due to binding with the correct or incorrect target, respectively, such that $\Delta G_{int,A} > \Delta G_{int,B}$. The correct binding probability is a sharp sigmoid that reaches half its maximal value at $\Delta G_{def} \approx \Delta G_{int,A}$. The incorrect binding probability decreases more steeply than the correct one and reaches half its maximal value at $\Delta G_{def} \approx \Delta G_{int,B}$. Thus, the cost function exhibits a minimum at a non-zero value, $\Delta G_{def}$ = 1/2( $\Delta G_{int,A} + \Delta G_{int,B}$). This implies that the recognizer and its target should not have complementary structures and thus conformational changes should occur upon binding. (b) The effect of the tolerance on the cost. In a system with low error-tolerance, $t > 1$, the penalty for binding a incorrect target is higher than the gain from binding a correct target. Small deformation is less beneficial since it allows more incorrect binding and thus the optimal deformation free energy is slightly shifted to larger values relative to $t$ = 1. Similarly, the opposite phenomenon occurs at high error-tolerance, $t < 1$, and the deformation free energy is shifted to lower values. In the figure $\Delta G_{int,B}/\Delta G_{int,A}$ = 1/2.

The effect of tolerance on the cost $C$ is shown in Fig. 2(b). The tendency of a system with low error-tolerance, $t > 1$, is to minimize the incorrect binding at the expense of correct binding. In such a system, the penalty for binding an incorrect target is higher than the gain from binding a correct target. A small deformation is less beneficial since it allows more incorrect binding and therefore the optimal deformation free energy is shifted to larger values relative to $t$ = 1. Similarly, the opposite phenomenon occurs at high error-tolerance, $t < 1$, and the deformation free energy is shifted to lower values.

## III. Optimal Conformational Mismatch and Flexibility

So far, we have discussed the optimal deformation free energy without accounting for the physical properties of the molecules such as conformation and flexibility. In order to evaluate the role of these parameters, their effect on the deformation energy must be specified. Modeling proteins as elastic networks was previously applied to study the fluctuations of proteins and to predict domain deformation upon binding [15-18].

Motivated by these studies, we treat the molecules as elastic networks and assume that the deformation energy is $E = (1/2)kd^2$, where $k$ is an effective spring constant and $d$ is the structural *mismatch* between the native structures of the molecules. The mismatch, $d$, may be a length difference if the deformation is an extension or an angle difference if the deformation is a bending. Thus, the distribution of conformations is a Gaussian centered around the native conformation of the molecules with a variance $\sigma \sim 1/k^{1/2}$. Once the binding sites are aligned, binding interaction energy $E_{int}$ is gained.

Using straightforward statistical mechanics calculation [11], the cost function (4) can now be expressed as

$$C = \frac{-1}{1+s\cdot e^{-E_{int,A}}\dfrac{1}{\sqrt{\bar{k}_A}}e^{(\bar{k}_A/2)d^2}} + t\cdot\frac{1}{1+s\cdot e^{-E_{int,B}}\dfrac{1}{\sqrt{\bar{k}_B}}e^{(\bar{k}_B/2)d^2}}, \qquad (5)$$

where $\bar{k}_{A/B}$ is the harmonic mean of the recognizer and the target spring constant, $\bar{k}_{A/B} = k_a k_{A/B} / (k_a + k_{A/B})$ . $E_{int,A}$ and $E_{int,B}$ are the binding energy gains due to interactions between the recognizer and the correct or incorrect target, respectively, such that $E_{int,A} > E_{int,B}$. The parameter $s$ is a measure for the kinetic phase space of the system [11] and does not depend on the structural parameters of the molecules. We can now minimize $C$ (5) and obtain the structural parameters that optimize detection [Fig. 3]. Our analysis shows that there are two optimal design "phases". In one phase, the native states of the optimal recognizer and the correct target are complementary, that is, the optimal mismatch is zero, $d = 0$. In the other phase, the native states are not complementary, $d \neq 0$, and, thus, conformational changes upon binding are beneficial.

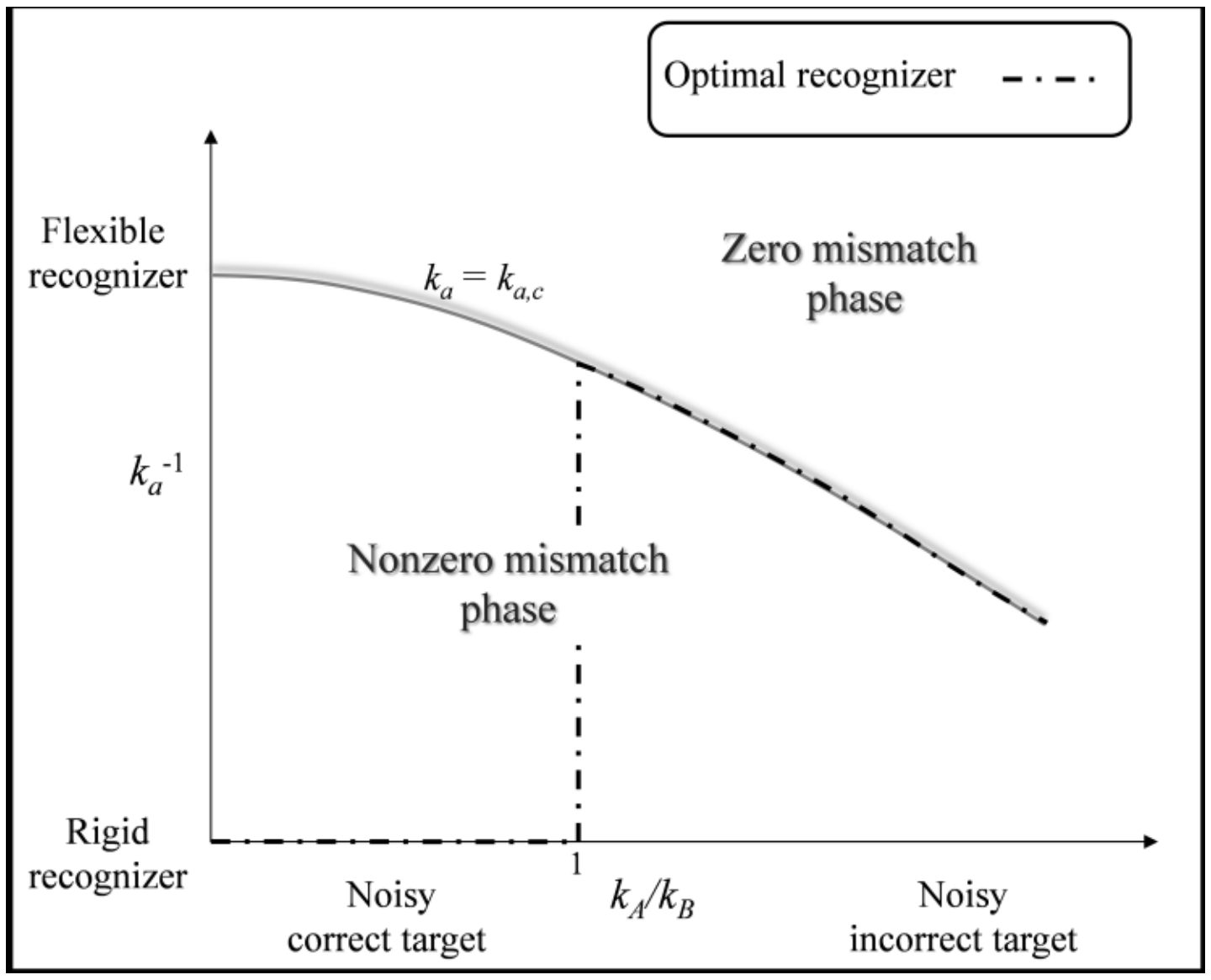

Fig. 3. Optimal design phases. There are two design phases for the optimal recognizer. In one, the native state structure of the recognizer should be complementary to the native state structure of its correct target. In the other, there is a structural mismatch between the native states and thus conformational changes occur upon binding. The optimal mismatch depends on the flexibility of the recognizer and its targets. For targets with similar flexibility, $k_A/k_B = 1$, the design depends on whether the spring constant of the recognizer is above or below the critical one, $k_{a,c}$. For a noisy incorrect target, $k_A/k_B > 1$, the optimal design is a flexible recognizer with $k_a \approx k_{a,c}$ and a zero mismatch. For a noisy correct target, $k_A/k_B < 1$, the optimal design is a rigid recognizer with a non-zero mismatch. Typical biological flexibilities correspond to $k_a \geq k_{a,c}$ and thus both designs may be beneficial.

If the targets have similar flexibility, $k_A = k_B = k_t$, as the flexibility of the recognizer, $k_a$, is varied, the system undergoes a phase transition between the zero mismatch and the non-zero mismatch phases [11]. In the symmetric case, $t$ =1, the critical spring constant at this transition is,

$$k_{a,c} = \frac{k_c k_t}{k_t - k_c}, \quad (6)$$

where $k_c = s^2 exp(E_{int,A} + E_{int,B})$. For a very flexible recognizer, $k_a < k_{a,c}$, the deformation energy is much smaller than the binding energy gain of both targets. Thus, introducing a mismatch does not provide any benefit. This is also true if the targets are very flexible, $k_t < k_c$. However, above the critical flexibility, $k_a > k_{a,c}$, the reconizer is more rigid and thus deformation can occur upon binding to the correct target, while it is not likely to occur upon binding to the incorrect target. Thus, the optimal mismatch has a nonzero value,

$$d_{opt} = \sqrt{\frac{\log(\bar{k} / k_c)}{\bar{k}}}. \quad (7)$$

As the recognizer becomes more rigid, the optimal mismatch decreases. Yet, the optimal mismatch for a rigid recognizer is still nonzero, $d_{opt}(k_a \rightarrow \infty) = (\log(k_t/k_c)/k_t)^{1/2}$.

When the tolerance is asymmetric, $t \neq 1$, the system still undergoes a phase transition but the values of the critical parameters change. As the tolerance of the systems to errors is reduced ($t > 1$), avoiding a wrong decision is more beneficial than making the correct one. As a result, the critical spring constant is lower. For a system with high tolerance ($t < 1$), the priority is the formation of a correct product and thus the critical spring constant is higher.

When the incorrect target is more flexible, that is fluctuates more, than the correct one, $k_A > k_B$, the ensemble of incorrect target conformations is more "spread" than the correct ensemble. The optimal design is a flexible recognizer, $k_a \approx k_{a,c}$, with zero mismatch that can sample many correct conformations while sampling only few incorrect ones. In other words, a stochastic decision unit will perform better than a deterministic one. In the case where the correct target is noisier, $k_A < k_B$, the ensemble of correct target conformations is more spread than the incorrect ensemble. Thus, the optimal design is a rigid recognizer with a nonzero mismatch relative to the main target.

## IV. Discussion

By applying the framework of signal detection to molecular recognition, we gain a quantitative insight on the optimal design of molecular recognizers in a noisy biochemical environment. The "phases" of optimal design depend on whether the recognizer flexibility is above or below a critical one [Fig. 3]. Typical structural parameters and interactions of biomolecules indicate that both design strategies, with or without a mismatch, are relevant for molecular recognition scenarios. If the competitor is noisier than the correct target, the optimal design is a recognizer with a critical flexibility and a zero mismatch. In this case, a noisy recognizer samples the target more efficiently than a deterministic, rigid, recognizer. If the correct target is noisier, the optimal design is a rigid recognizer with a non-zero mismatch which is similar to a rigid key that does not fit exactly into its lock.

The result that conformational changes may provide optimal recognition, may explain their abundance in nature as a mechanism that increases the fitness of the recognition process. We thus suggest a general design principle, termed *conformational proofreading*, in which the introduction of a structural mismatch between the recognizer and its target enhances the quality of detection. Besides rationalizing observed natural processes, this kind of formalism may be used in the design of future synthetic biological recognition systems.

More realistic and empirical examples can be incorporated into this framework in order to evaluate their optimal design. Such an example is the process of homologues recombination in which homologous DNA segments can be exchanged. Correct homologous recombination requires the detection of a specific homologous DNA sequence within a huge variety of heterologous sequences. During this process it is known that one of the DNA segments is deformed leading to a large mismatch relative to the target DNA. We suggest that this deformation can be explained using signal detection approach [12].